# *Super collimation of the radiation by a point source in a uniaxial wire medium*

Tiago A. Morgado, Mário G. Silveirinha[*]

*Department of Electrical Engineering, Instituto de Telecomunicações, University of Coimbra, 3030 Coimbra, Portugal*

*E-mail:* tiago.morgado@co.it.pt, mario.silveirinha@co.it.pt

## Abstract

We investigate the radiation properties of a short horizontal dipole embedded in a uniaxial wire medium. It is shown that the uniaxial wire medium enables a super-collimation of the dipole radiation such that the radiation pattern has a singularity and the radiated fields are non-diffractive in the broadside direction. We derive a closed analytical formula for the power radiated by the dipole. Our theory demonstrates that as a consequence of the ultrahigh density of photonic states of the nanowire array, the power radiated by the dipole is strongly enhanced as compared to that emitted in the dielectric host material.

---

[*] To whom correspondence should be addressed: E-mail: mario.silveirinha@co.it.pt

# I. INTRODUCTION

The uniaxial wire medium - a periodic array of parallel metallic wires embedded in a dielectric host – is one of the most important and extensively studied metamaterials [1-9]. Nanowire materials have attracted considerable attention due to their peculiar electromagnetic properties, such as the strongly spatially dispersive (nonlocal) response [4-9] and an anomalously high density of photonic states [10-14]. These unusual properties are useful in many applications from microwave up to optical frequencies [9-26].

Numerous theoretical and analytical methods were developed in the last decade to accurately characterize the effective electromagnetic response of wire media [3-8]. Such tools make possible, for instance, the study of the wave propagation in wire medium slabs [7, 27-29] and solving source-free spectral problems for the natural modes in closed analytical form [30-32]. However, the study of the problem of radiation by localized sources embedded in a wire medium background was somehow on the back burner for a long time. Only recently this subject has been investigated in more detail [33-36]. In particular, the radiation properties of a short vertical dipole embedded in a uniaxial wire medium were investigated in Ref. [35], using both a nonlocal framework [4, 6] and a quasi-static approach relying on additional variables that describe the internal degrees of freedom of the medium [8]. The objective of this work is to further study the radiation problem of a short dipole embedded in a uniaxial wire medium. Specifically, we extend the analysis to the scenario wherein the dipole is horizontal with respect to the wires [see Fig. 1] rather than vertical as in Ref. [35]. Importantly, we demonstrate that for a horizontal dipole the radiated fields can be super-collimated by the nanowires leading to a super-directive emission along the axial direction.

This paper is organized as follows. In Sec. II, we introduce the radiation problem under study. In Sec. III, we present the solution of the problem in the spectral domain using the nonlocal dielectric function approach. In Sec. IV, we derive a closed analytical formula for the power radiated by the short dipole in the metamaterial. Finally, in Sec. V the conclusions are drawn. Throughout this work we assume a time harmonic regime with time dependence of the form $e^{-i\omega t}$.

## II. THE RADIATION PROBLEM

Figure 1 illustrates the geometry of the problem under study: a square array of metallic wires embedded in a dielectric host. The spacing between the wires is $a$ and the wire radius is $r_w$. The excitation source is centered at the position $\mathbf{r}' = (x', y', z')$ and corresponds to a short horizontal dipole described by the electric current density $\mathbf{j}_{\text{ext}}(\mathbf{r}) = -i\omega p_e \delta(\mathbf{r} - \mathbf{r}')\hat{\mathbf{x}}$, where $\omega$ is the angular frequency of oscillation, $p_e$ represents the electric dipole moment, and $\hat{\mathbf{x}}$ is the unit vector along the positive *x*-axis. Our objective is to characterize the radiation emitted by the short dipole using an effective medium theory.

As a starting point, we remark that the use of effective medium methods requires that the source must be localized in a region with characteristic dimensions larger than the lattice constant $a$ of the metamaterial [see Fig. 1], so that it is possible to assume that only waves with $-\pi/a \leq k_x, k_y \leq \pi/a$ can be excited in the metamaterial [35]. This property can be justified by the "uncertainty principle" of the Fourier transform, which establishes that the spreading of a function in the spatial and spectral domains is not independent, and the characteristic widths in the spatial ($\sigma_x$) and spectral ($\sigma_{k_x}$) domains are bound to satisfy $\sigma_x \sigma_{k_x} \geq \frac{1}{2}$. Thus, if the characteristic width of a source is of the order of $\sigma_x \sim a$, one may assume $\sigma_{k_x} \sim 1/a$, which justifies taking a wave vector cut-off of the order $k_{\max} \sim \pi/a$. Thus, even though we represent the dipole using the Dirac distribution $\delta$-symbol, in practice the dipole needs to have length/radius at least comparable to *a*, and should at the same time be much shorter than the wavelength. Such a current distribution can be modeled mathematically by $\delta(\mathbf{r}) = \frac{1}{(2\pi)^3} \int_{-k_{max}}^{k_{max}} dk_x \int_{-k_{max}}^{k_{max}} dk_y \int_{-\infty}^{+\infty} dk_z e^{i\mathbf{k}\cdot\mathbf{r}}$ being $k_{max}$ the spatial cut-off. Note that we do not need a spatial cut-off along the z-direction because the structure is invariant to translations along z. For $k_{max} \to \infty$ we recover the usual Dirac distribution.

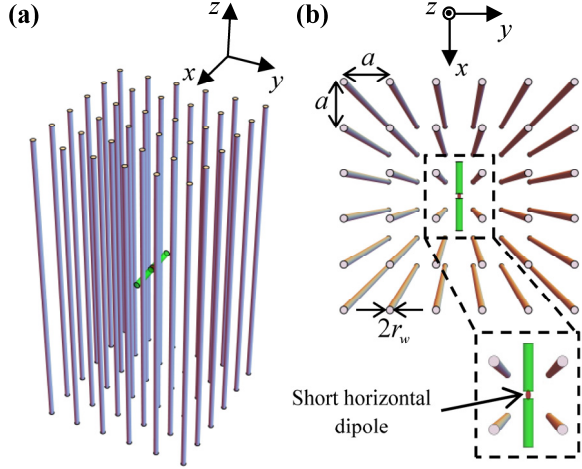

**Fig. 1.** Illustration of the system under study: a short horizontal dipole is embedded in an unbounded uniaxial wire medium formed by metallic wires embedded in a dielectric host. The wires have radius $r_w$ and are arranged in a square lattice with lattice constant $a$. (a) Perspective view; (b) Top view together with a zoom-in view showing the area surrounding the short dipole.

It is known [4, 6] that the uniaxial wire medium formed by straight metallic wires is characterized by the effective dielectric function

$$\frac{\bar{\bar{\varepsilon}}(\omega,\mathbf{k})}{\varepsilon_0 \varepsilon_h} = \bar{\bar{I}} + \chi_{zz}(\omega,k_z)\hat{\mathbf{z}} \otimes \hat{\mathbf{z}}, \qquad (1)$$

where $\otimes$ denotes the tensor product of two vectors and $\varepsilon_h$ is the relative permittivity of the host material. In case of perfect electrical conducting (PEC) wires the $zz$ susceptibility is given by $\chi_{zz} = -\frac{k_p^2}{k_h^2 - k_z^2}$ where $k_p$ is the plasma wave number of the wire medium [4, 6], and $k_h = k_0\sqrt{\varepsilon_h} = \omega\sqrt{\mu_0 \varepsilon_0 \varepsilon_h}$ is the wave number in the host region. The formula for $\chi_{zz}$ in the general case of lossy metallic wires can be easily obtained from Ref. [6]. Note that the effective dielectric function depends explicitly on $k_z \leftrightarrow -i\partial/\partial z$.

Next, based on this nonlocal framework, we calculate the fields radiated by the horizontal dipole in the unbounded uniaxial wire medium and derive an explicit formula for the radiated power.

## III. THE RADIATED FIELDS

The strategy to determine the emitted fields is to solve the radiation problem in the spectral domain (i.e., in the Fourier spatial domain) and then, calculate the inverse Fourier transform to obtain the field distributions in the spatial domain.

It is known [35] that in a spatially dispersive medium, the Maxwell equations may be written in the space domain as follows:

$$\nabla \times \mathbf{E} = i\omega\mu_0 \mathbf{H} \tag{2}$$

$$\nabla \times \mathbf{H} = -i\omega\overline{\overline{\varepsilon}}(\omega,-i\nabla)\cdot \mathbf{E} + \mathbf{j}_{\text{ext}}. \tag{3}$$

The dyadic operator $\overline{\overline{\varepsilon}}(\omega,-i\nabla)$ represents the effective dielectric function of the material and can be written explicitly as a convolution. In the spectral (Fourier) domain, we have the correspondence $-i\nabla \leftrightarrow \mathbf{k} = (k_x, k_y, k_z)$. Hence, after some straightforward manipulations of Eqs. (2)-(3), one can find that the Fourier transform of the electric field satisfies [35]

$$\mathbf{E}(\omega,\mathbf{k}) = -i\omega\mu_0 \left[\omega^2\mu_0\overline{\overline{\varepsilon}}(\omega,\mathbf{k}) + \mathbf{k}\otimes\mathbf{k} - k^2\overline{\overline{\mathbf{I}}}\right]^{-1}\cdot \mathbf{j}_{\text{ext}}. \tag{4}$$

with $\mathbf{j}_{\text{ext}}(\mathbf{k}) = -i\omega p_e \hat{\mathbf{x}}$. From Eq. (4), it follows that:

$$\mathbf{E}(\omega,k_\parallel,k_z) = \frac{p_e}{\varepsilon_0\varepsilon_h}\frac{1}{k^2-k_h^2}\left[\left(k_h^2 - k_x^2 \frac{k_h^2 - k^2 + \chi_{zz}k_h^2}{k_h^2 - k^2 + \chi_{zz}(k_h^2 - k_z^2)}\right)\hat{\mathbf{x}} \right. \\ \left. -\left(\frac{k_h^2 - k^2 + \chi_{zz}k_h^2}{k_h^2 - k^2 + \chi_{zz}(k_h^2 - k_z^2)}\right)k_xk_y\hat{\mathbf{y}} + \frac{k^2 - k_h^2}{k_h^2 - k^2 + \chi_{zz}(k_h^2 - k_z^2)}k_xk_z\hat{\mathbf{z}}\right], \tag{5}$$

where $k_\parallel = \sqrt{k_x^2 + k_y^2}$ and $k^2 = k_x^2 + k_y^2 + k_z^2$. For completeness, we mention that the radiated fields can also be written in terms of an electric Hertz potential $\mathbf{\Pi}_e$ in the usual manner:

$$\mathbf{E}(\omega,\mathbf{k}) = \frac{\omega^2}{c^2}\varepsilon_h \mathbf{\Pi}_e - \mathbf{k}(\mathbf{k}\cdot\mathbf{\Pi}_e), \tag{6}$$

$$\mathbf{H}(\omega,\mathbf{k}) = \omega\varepsilon_0\varepsilon_h \mathbf{k}\times\mathbf{\Pi}_e, \tag{7}$$

with

$$\mathbf{\Pi}_e(\omega,\mathbf{k}) = \frac{1}{k^2 - k_h^2}\frac{p_e}{\varepsilon_0\varepsilon_h}\left(\hat{\mathbf{x}} + \hat{\mathbf{z}}\frac{\chi_{zz}k_x k_z}{k_h^2 - k^2 + \chi_{zz}\left(k_h^2 - k_z^2\right)}\right). \tag{8}$$

In particular, the *x*-component of the Hertz potential can be explicitly integrated so that $\Pi_{e,x}(\omega,\mathbf{r}) = \frac{p_e}{\varepsilon_0\varepsilon_h}\frac{e^{ik_h r}}{4\pi r}$ with $r = \sqrt{x^2 + y^2 + z^2}$. Yet, for the analytical developments of the next section it is more useful to work directly with Eq. (5).

## A. PEC nanowires

In the following, we focus our analysis on the particular case wherein the nanowires are PEC. In such a case, we get from Eq. (5):

$$\mathbf{E}(\omega,k_\parallel,k_z) = \frac{p_e}{\varepsilon_0\varepsilon_h}\left[\hat{\mathbf{x}}\left(-\frac{k_h^2 k_p^2 k_x^2}{k_\parallel^2\left(k_p^2 + k_\parallel^2\right)}\frac{1}{k_h^2 - k_z^2} + k_h^2\left(1 - \frac{k_x^2}{k_\parallel^2}\right)\frac{1}{k_\parallel^2 + k_z^2 - k_h^2}\right.\right.$$
$$\left. + \frac{\left(k_h^2 - k_p^2 - k_\parallel^2\right)k_x^2}{k_p^2 + k_\parallel^2}\frac{1}{k_p^2 + k_\parallel^2 + k_z^2 - k_h^2}\right)$$
$$+ \hat{\mathbf{y}}\left(-\frac{k_h^2 k_p^2 k_x k_y}{k_\parallel^2\left(k_p^2 + k_\parallel^2\right)}\frac{1}{k_h^2 - k_z^2} - \frac{k_h^2 k_x k_y}{k_\parallel^2}\frac{1}{k_\parallel^2 + k_z^2 - k_h^2}\right.$$
$$\left. + \frac{\left(k_h^2 - k_p^2 - k_\parallel^2\right)k_x k_y}{k_p^2 + k_\parallel^2}\frac{1}{k_p^2 + k_\parallel^2 + k_z^2 - k_h^2}\right)$$
$$\left. - \hat{\mathbf{z}}\frac{k_x k_z}{\left(k_p^2 + k_\parallel^2 + k_z^2 - k_h^2\right)}\right], \tag{9}$$

Interestingly, it is possible to calculate explicitly the inverse Fourier transform of the electric field in $k_z$. A straightforward analysis shows that the components of the electric field satisfy:

$$E_x(\omega, \mathbf{k}_\parallel, z) = \frac{1}{2}\frac{p_e}{\varepsilon_0 \varepsilon_h}\left(i\frac{k_h k_p^2 k_x^2}{k_\parallel^2(k_p^2 + k_\parallel^2)}e^{ik_h|z|} + \frac{k_h^2}{\gamma_h}\left(1 - \frac{k_x^2}{k_\parallel^2}\right)e^{-\gamma_h|z|} - \frac{k_x^2 \gamma_{TM}}{k_p^2 + k_\parallel^2}e^{-\gamma_{TM}|z|}\right), \quad (10)$$

$$E_y(\omega, \mathbf{k}_\parallel, z) = \frac{1}{2}\frac{p_e}{\varepsilon_0 \varepsilon_h}\left(i\frac{k_h k_p^2 k_x k_y}{k_\parallel^2(k_p^2 + k_\parallel^2)}e^{ik_h|z|} - \frac{k_h^2}{\gamma_h}\frac{k_x k_y}{k_\parallel^2}e^{-\gamma_h|z|} - \frac{k_x k_y \gamma_{TM}}{k_p^2 + k_\parallel^2}e^{-\gamma_{TM}|z|}\right), \quad (11)$$

$$E_z(\omega, \mathbf{k}_\parallel, z) = -i\frac{p_e}{\varepsilon_0 \varepsilon_h}k_x \frac{\text{sgn}(z)}{2}e^{-\gamma_{TM}|z|}. \quad (12)$$

where $\gamma_{TM} = \sqrt{k_p^2 + k_\parallel^2 - k_h^2}$ is the $z$-propagation constant of the transverse magnetic (TM) waves that propagate in the uniaxial wire medium, and $\gamma_h = \sqrt{k_\parallel^2 - k_h^2}$ is the $z$-propagation constant of the transverse electric (TE) waves. It is worth noting that each term of Eqs. (10-11) is related to the contribution of each of the three eigenmodes supported by the uniaxial wire medium [4, 6, 8], namely the transverse electromagnetic (TEM) mode (term associated with the $e^{ik_h|z|}$ spatial variation), the transverse electric (TE) mode (term associated with the $e^{-\gamma_h|z|}$ spatial variation), and the transverse magnetic (TM) mode (term associated with the $e^{-\gamma_{TM}|z|}$ spatial variation).

The most important radiation channel is associated with the TEM waves which have the propagation factor $e^{ik_h|z|}$. The corresponding radiated fields can be explicitly integrated, and can be written in terms of an electric potential $\phi$ as follows:

$$\mathbf{E}^{TEM}(\omega, \mathbf{r}) = -e^{ik_h|z|}\nabla_t \phi, \quad (13)$$

with $\nabla_t = \frac{\partial}{\partial x}\hat{\mathbf{x}} + \frac{\partial}{\partial y}\hat{\mathbf{y}}$ and

$$\phi(\omega, \mathbf{r}) = -\frac{p_e}{2\varepsilon_0 \varepsilon_h}ik_h \frac{x}{\rho}\frac{1 - k_p \rho K_1(k_p \rho)}{2\pi\rho}. \quad (14)$$

where $\rho = \sqrt{x^2 + y^2}$, and $K_n$ represents the modified Bessel function of the second kind and order $n$. This result is rather interesting, since it predicts that a non-diffractive cylindrical beam extremely confined to the wires axis can be excited inside the nanowire metamaterial. Note that the TEM beam does not decay with the distance as it propagates away from the point source, different from what happens in a standard dielectric wherein the fields decay as $1/r$. As a consequence, the radiation intensity of the horizontal dipole diverges to infinity for an observation direction parallel to the $z$-direction. Thus, the beam is super-collimated by the nanowires and the directivity of this elementary radiator diverges to infinity. It should be noted that the described behavior is fundamentally different from that found in Ref. [35] for a vertical dipole. Indeed, a vertical dipole is unable to excite the TEM waves in the nanowire material, and thus such an excitation cannot generate a diffractionless beam.

The total radiated field has also contributions from the excited TE and TM waves. For long wavelengths, $k_h \ll k_p$, the contribution of the TM mode is negligible in the far-field region because its attenuation constant is very large [29]. On the other hand, the TE waves lead to spherical wavefronts, but it will be shown later that they only transport a small fraction of the total radiated power, and hence in practice they are of secondary importance. Moreover, it can be analytically shown that in the $xoz$ plane the electric field associated with the TE mode decays as $1/r^2$ along any observation direction, with the exception of the $z$-axis.

Notably, it may be verified that the field $\mathbf{E}^{\text{TEM}}$ is singular along the $z$-axis such that $E_x$ diverges logarithmically as $\rho \to 0$. This unphysical behavior is due to the fact that in the continuous limit (when $-\infty < k_x, k_y < \infty$) the density of photonic states of the wire medium diverges [20, 22], and hence the metamaterial has infinite radiation channels leading to a spatial singularity of the radiated field. This problem can be easily fixed by introducing the spatial cut-off $k_{\max} = \pi / a$ such that the integration range in $k_x$ and $k_y$ is truncated to the first

Brillouin zone, $-\pi/a \leq k_x, k_y \leq \pi/a$. Indeed, the wave vector of the TEM modes must be restricted to the first Brillouin zone when the actual granularity of the metamaterial is properly considered [20]. The truncation of $k_x$ and $k_y$ to the first Brillouin zone is also consistent with our assumption that the point source is less localized than the period of the wire medium.

### B. Numerical example

To illustrate the non-diffractive nature of the radiation transported by the TEM waves, we represent in Fig. 2 the *x*-component of $\mathbf{E}^{\text{TEM}}$ in the $y = 0$ plane (the E-plane) and in the $x = 0$ plane (the H-plane). As seen, the field is strongly confined to the *z*-axis and is guided away from the source without suffering any lateral spreading (no diffraction). This result is a consequence of the channeling properties of the uniaxial wire medium [15-18], which enable collimating the near field of a source.

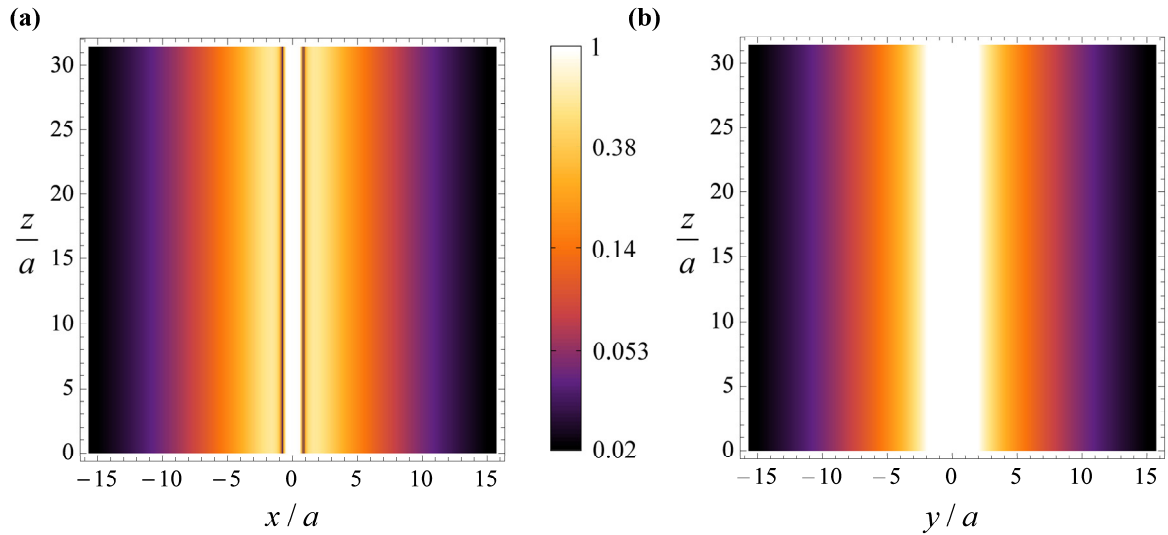

**Fig. 2.** Density plot of $\left|E_x^{\text{TEM}}\right|$ (a) in the $y = 0$ plane (the E-plane) and (b) in the $x = 0$ plane (the H-plane), obtained from Eqs. (13)-(14). The lattice period is $a$, the radius of the wires is $r_w = 0.01a$, the host is a vacuum $\varepsilon_h = 1$, and the frequency of operation is such that $\omega a/c = 0.1$. The dipole is centered at the origin ($(x, y, z) = (0, 0, 0)$).

We also numerically computed the total radiated fields in the E-plane by numerically calculating the inverse Fourier transform of Eqs. (10)-(12):

$$\mathbf{E}(\mathbf{r},\omega) = \frac{1}{(2\pi)^2} \iint \mathbf{E}(\omega,\mathbf{k}_\parallel,z) e^{i(k_x x + k_y y)} dk_x dk_y. \tag{15}$$

To have a direct comparison with Eqs. (13)-(14), in a first stage the integration range is taken $-\infty < k_x, k_y < \infty$, i.e. similar to the previous subsection the spatial cut-off is neglected ($k_{max} = \infty$). It can be checked that the only nonzero components of the electric field in the E-plane are $E_x$ and $E_z$. The $E_z$ component of the fields can be explicitly integrated and satisfies:

$$E_z(\omega,\mathbf{r}) = \frac{p_e}{\varepsilon_0 \varepsilon_h} \frac{\partial}{\partial x} \frac{\partial}{\partial z} \left( \frac{1}{4\pi r} e^{i k_{ef} r} \right). \tag{16}$$

where $k_{ef} = \sqrt{k_h^2 - k_p^2}$. In addition, we have also calculated the emitted fields using the wave vector cut-off $-\pi/a \leq k_x, k_y \leq \pi/a$, so that the integration range in Eq. (15) is restricted to the first Brillouin zone.

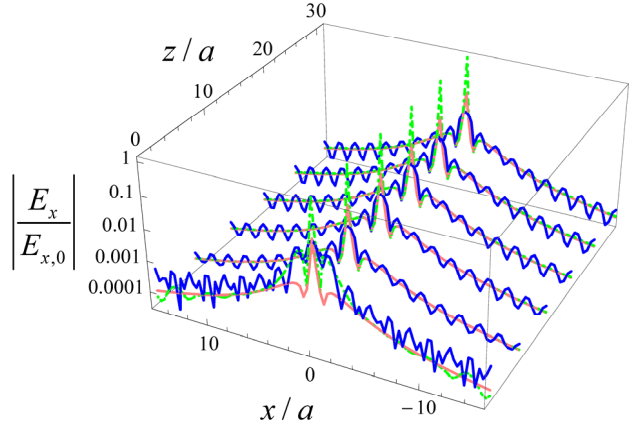

**Fig. 3.** Field profile of $E_x$ as a function of $x$ and for different positions along $z$-direction. Pink solid curves: contribution of the TEM mode with no spatial cut-off [Eq. (13)]; green dot-dashed curves: numerical result calculated with Eq. (15) with the integration range $\{k_x, k_y\} \in (-\infty, +\infty)$; blue solid curves: numerical result calculated with Eq. (15) with the integration range $\{k_x, k_y\} \in [-\pi/a, \pi/a]$. The structural parameters are the same as in Fig. 2.

In Fig. 3 we depict the field profiles of $E_x$ for different positions along the $z$-direction, calculated by numerically integrating Eq. (15) with no spatial cut-off (green dot-dashed curves) and with a spatial cut-off (blue solid curves). Figure 3 also shows the individual contribution of the TEM modes (without a spatial cut-off) given by the analytical formula (13) (pink solid curves). As seen, there is an overall good agreement between the three calculation methods, especially for the analytical and the numerical results calculated with no spatial cut-off. Only in the very near field some differences are discernible. This result confirms that the TEM waves determine, indeed, the main emission channel. In particular, the results show a strong confinement of the dipole radiation to the wires axis and the guiding of the emitted fields with no diffraction. As expected, with a spatial cut-off the field singularity along the z-axis disappears. Interestingly, the spatial cut-off also introduces some ripple in the field profiles due to the spatial averaging of the field singularities (in the absence of spatial cut-off) along the z-axis. It is important to highlight that imposing a wave vector cut-off is equivalent to perform a low-pass spatial filtering.

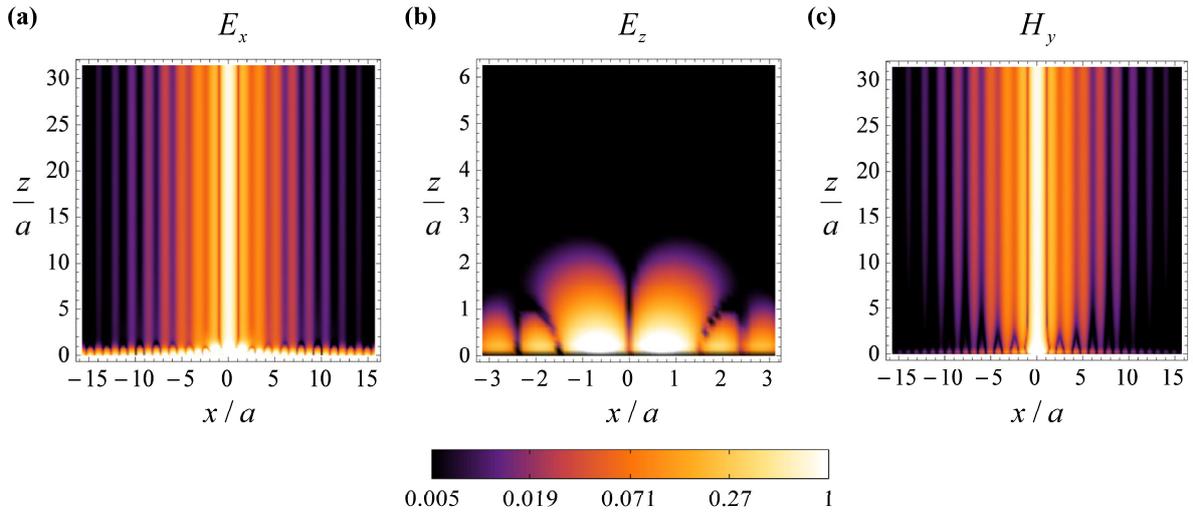

**Fig. 4.** Density plot of the electromagnetic fields amplitude in the $y=0$ plane (E-plane), calculated with Eq. (15) with the spatial cut-off $\{k_x, k_y\} \in [-\pi/a, \pi/a]$. (a) $|E_x|$; (b) $|E_z|$; (c) $|H_y|$ for the same structural parameters as in Fig. 2.

Figure 4 shows the exact spatial distribution of the electromagnetic field ($E_x$, $E_z$, and $H_y$) in the E-plane obtained with Eq. (15) using a spatial cut-off. The y-component of the magnetic field is determined from the Maxwell's equations, $H_y = \frac{i}{k_0 \eta_0}\left(\frac{\partial}{\partial x}E_z - \frac{\partial}{\partial z}E_x\right)$, where $\eta_0$ is the free-space impedance. The spatial derivatives are calculated using finite differences [38]. The result of Fig. 4(b) demonstrates that for $k_h \ll k_p$, the contribution of the TM mode is negligible in the far field. Note that the $E_z$ component depends exclusively on the TM modes, and is confined to the near-field region. In contrast, similar to the $E_x$ component, the $H_y$ field is also characterized by a diffraction-free beam pattern in the far field (see Fig. 4(c)).

## IV. THE RADIATED POWER

Next, relying on the eigenwave expansion formalism introduced in Ref. [37], we derive a closed analytical formula for the power radiated by the short horizontal dipole inside the unbounded uniaxial wire medium [see Fig. 1]. To begin with, we present an overview of the eigenfunction expansion formalism. Then, we use this formalism to calculate the radiated power for PEC nanowires.

### A. Overview of the eigenfunction expansion formalism

In Ref. [37], we derived a general analytical formulation that enables calculating the power emitted by moving sources in frequency dispersive lossless wire media. This formalism can be applied in a straightforward manner to standard non-moving sources, e.g. to an electric dipole. Specifically, using Eq. (11) of Ref. [37] and $\mathbf{j}_{\text{ext}} = -i\omega p_e \delta(\mathbf{r} - \mathbf{r}')\hat{\mathbf{x}}$ it is simple to check that the electric field $\mathbf{E}$ radiated by the point dipole in time-harmonic regime has the following exact modal expansion:

$$\mathbf{E}(\mathbf{r}) = \frac{\omega p_e}{2V} \sum_n \frac{1}{\omega_n - \omega} \mathbf{E}_n(\mathbf{r}) \mathbf{E}_n^*(\mathbf{r}') \cdot \hat{\mathbf{x}} \ . \qquad (17)$$

Here $V = L_x \times L_y \times L_z$ is the volume of the region of interest, which in the end will be let to approach $V \to \infty$. The symbol "*" denotes complex conjugation. The summation in Eq. (17)

is over the electromagnetic (plane wave) modes of the bulk material $(\mathbf{E}_n, \mathbf{H}_n)$ $n=1,2,\ldots$, which for dispersive media must be normalized as follows [37]:

$$\frac{1}{2}\mathbf{E}_n^* \cdot \frac{\partial}{\partial \omega}\left[\omega \bar{\bar{\varepsilon}}(\omega,\mathbf{k})\right]_{\omega=\omega_n} \cdot \mathbf{E}_n + \frac{1}{2}\mu_0 \mathbf{H}_n^* \cdot \mathbf{H}_n = 1. \tag{18}$$

The frequencies $\omega_n$ are the real-valued eigenfrequencies of the natural modes. Importantly, the summation in Eq. (17) must include both the positive frequency and the negative frequency eigenmodes [37].

Since we are dealing with continuous media, it is clear that the eigenmodes $\mathbf{E}_{n,\mathbf{k}}$ can be taken as plane waves associated with a wave vector $\mathbf{k}$. Hence, Eq. (17) becomes

$$\mathbf{E}(\mathbf{r}) = \frac{\omega p_e}{2V}\sum_n \sum_\mathbf{k} \frac{1}{\omega_{n,\mathbf{k}} - \omega} \mathbf{E}_{n,\mathbf{k}}(\mathbf{r})\mathbf{E}_{n,\mathbf{k}}^*(\mathbf{r}') \cdot \hat{\mathbf{x}}, \tag{19}$$

where $\omega_{n,\mathbf{k}}$ are the natural frequencies associated with the plane waves $\mathbf{E}_{n,\mathbf{k}}$ with wave vector $\mathbf{k}$ and index $n$ ($n$ determines the eigenmode type). As discussed in Ref. [37], the eigenmodes $\mathbf{E}_n$ can be divided in four different types: TE modes, TM modes, TEM modes, and longitudinal (LS) (electrostatic and magnetostatic) modes with $\omega_{n,\mathbf{k}} = 0$. The LS modes do not contribute to the radiation field. However, they are mathematically important, since one cannot obtain a complete set of eigenfunctions without them [42].

In the continuous limit ($V \to \infty$), the summation over $\mathbf{k}$ is replaced by an integral and Eq. (19) becomes:

$$\mathbf{E}(\mathbf{r}) = \frac{\omega p_e}{16\pi^3}\sum_n \int d^3\mathbf{k} \frac{1}{\omega_{n,\mathbf{k}} - \omega} \mathbf{E}_{n,\mathbf{k}}(\mathbf{r})\mathbf{E}_{n,\mathbf{k}}^*(\mathbf{r}') \cdot \hat{\mathbf{x}}. \tag{20}$$

### B. The emitted power

The integrand of Eq. (20) is singular when $\omega_{n,\mathbf{k}} = \omega$, which corresponds to the isofrequency contours of the eigenmodes. Then, given that

$d^3\mathbf{k} = ds(\mathbf{k})dk_\perp = ds(\mathbf{k})d\omega/|\nabla_\mathbf{k}\omega_{n,\mathbf{k}}|$ ($ds(\mathbf{k})$ is the element of area of the isofrequency surfaces), one can write Eq. (20) as follows:

$$\mathbf{E}(\mathbf{r}) = \frac{\omega p_e}{16\pi^3}\sum_n \int ds(\mathbf{k}) \int d\omega \frac{1}{\omega_{n,\mathbf{k}}-\omega}\frac{1}{|\nabla_\mathbf{k}\omega_{n,\mathbf{k}}|}\mathbf{E}_{n,\mathbf{k}}(\mathbf{r})\mathbf{E}^*_{n,\mathbf{k}}(\mathbf{r}')\cdot\hat{\mathbf{x}}. \tag{21}$$

To avoid the singularity of the integrand we replace $\omega \to \omega + i0^+$, such that the integration path is in the upper-half frequency plane, consistent with the causality of the system response [37]. Then using the identity $\frac{1}{x-i0^+} = \text{P.V.}\frac{1}{x} + i\pi\delta(x)$ [42], where P.V. denotes the Cauchy principal value, we may write Eq. (21) as follows:

$$\mathbf{E}(\mathbf{r}) = \frac{\omega p_e}{16\pi^3}\sum_n \int ds(\mathbf{k})\,\text{P.V.}\left(\int d\omega \frac{1}{\omega_{n,\mathbf{k}}-\omega}\frac{1}{|\nabla_\mathbf{k}\omega_{n,\mathbf{k}}|}\mathbf{E}_{n,\mathbf{k}}(\mathbf{r})\mathbf{E}^*_{n,\mathbf{k}}(\mathbf{r}')\cdot\hat{\mathbf{x}}\right) +$$
$$+ i\frac{\omega p_e}{16\pi^2}\sum_n \int_{\omega_{n,\mathbf{k}}=\omega} ds(\mathbf{k})\frac{1}{|\nabla_\mathbf{k}\omega_{n,\mathbf{k}}|}\mathbf{E}_{n,\mathbf{k}}(\mathbf{r})\mathbf{E}^*_{n,\mathbf{k}}(\mathbf{r}')\cdot\hat{\mathbf{x}} \tag{22}$$

In a time-harmonic regime the time-averaged radiated power is given by $P_{\text{rad}} = -\frac{1}{2}\int d^3\mathbf{r}\,\text{Re}\{\mathbf{E}\cdot\mathbf{j}^*_{\text{ext}}\}$, being $\mathbf{E}$ the macroscopic electric field and $\mathbf{j}_{\text{ext}} = -i\omega p_e\delta(\mathbf{r}-\mathbf{r}')\hat{\mathbf{x}}$ the electric current density. The term associated with the principal value integral in Eq. (22) does not contribute to the radiated power, and hence $P_{\text{rad}}$ can be written simply as:

$$P_{\text{rad}} = \frac{\omega^2|p_e|^2}{32\pi^2}\sum_n \int_{\omega_{n,\mathbf{k}}=\omega} ds(\mathbf{k})\frac{|\mathbf{E}_{n,\mathbf{k}}(\mathbf{r}')\cdot\hat{\mathbf{x}}|^2}{|\nabla_\mathbf{k}\omega_{n,\mathbf{k}}|}. \tag{23}$$

### C. PEC nanowires

To illustrate the application of the theory, next we suppose that the metallic wires are perfect conductors. It is known that for long wavelengths ($k_h \ll k_p$) the only propagating modes in the uniaxial wire medium are the TE and the TEM modes [7, 12]. Hence, for long wavelengths the summation in Eq. (23) can be restricted to TE and TEM modes.

The dispersion characteristic of the (positive frequency) TEM eigenmodes is given by $\omega = |k_z| c_h$, where $c_h = c/\sqrt{\varepsilon_h}$ [7]. On the other hand, the electric field of the TEM modes is of the form $\mathbf{E}_{\text{TEM},\mathbf{k}} \sim A\mathbf{k}_\parallel$, wherein $A$ is a normalization constant determined by Eq. (18). It was proven in Ref. [37] that $A$ satisfies $A = k_p / \sqrt{\left(k_x^2 + k_y^2\right)\left(k_x^2 + k_y^2 + k_p^2\right)\varepsilon_0 \varepsilon_h}$. Substituting this result into Eq. (23), it is found that the radiated power associated to the TEM modes is given by:

$$P_{\text{rad,TEM}} = \frac{\omega^2 |p_e|^2}{32\pi^2} 2 \times \int_{-\pi/a}^{\pi/a} \int_{-\pi/a}^{\pi/a} \frac{k_x^2}{c_h} \frac{k_p^2}{\left(k_x^2 + k_y^2\right)\left(k_x^2 + k_y^2 + k_p^2\right)\varepsilon_0 \varepsilon_h} dk_x dk_y. \tag{24}$$

The integration range was restricted to the Brillouin zone because it is assumed that the point source is less localized than the lattice period $a$ ($k_{\max} = \pi/a$). Indeed, it is essential to include the spatial-cutoff in the calculation of the emitted power, otherwise it diverges. The leading factor of 2 follows from the fact that the isofrequency surfaces of the TEM waves are formed by two parallel planar sheets. For convenience, next we replace the integration over the square shaped Brillouin zone by an integration over a circular Brillouin zone with radius $K_{\max} = 2\sqrt{\pi}/a$, such that the integral (24) becomes:

$$P_{\text{rad,TEM}} = \frac{\omega^2 |p_e|^2}{16\pi^2} \frac{k_p^2}{c_h \varepsilon_0 \varepsilon_h} \int_0^{2\pi} \int_0^{K_{\max}} \frac{\cos^2 \varphi}{k^2 + k_p^2} k\, dk d\varphi. \tag{25}$$

Straightforward calculations show that:

$$P_{\text{rad,TEM}} = \frac{\omega^2 |p_e|^2 k_p^2}{32\pi c_h \varepsilon_0 \varepsilon_h} \ln\left(1 + \frac{4\pi}{a^2 k_p^2}\right). \tag{26}$$

Equation (26) shows that the power radiated by the dipole inside the uniaxial wire medium increases as the separation between the wires $a$ decreases. This can be understood as a consequence of the enhancement of the density of TEM modes when the distance between the wires becomes increasingly smaller. The density of photonic states for the TEM modes is

$$D_{\text{TEM}}(\omega) = \frac{1}{(2\pi)^3} \int_{\omega_{\text{TEM},\mathbf{k}}=\omega} ds(\mathbf{k}) \frac{1}{|\nabla_{\mathbf{k}} \omega_{\text{TEM},\mathbf{k}}|} = 1/(\pi a^2 c_h)$$ [10-11]. Note that from Eq. (18) we may estimate that $|\mathbf{E}_{n,\mathbf{k}}(\mathbf{r}') \cdot \hat{\mathbf{x}}|^2 \sim \frac{1}{\varepsilon_0 \varepsilon_h}$. Using this approximation in Eq. (23) we get $P_{\text{rad,TEM}} \approx \frac{\pi}{4} \frac{\omega^2 |p_e|^2}{\varepsilon_0 \varepsilon_h} D_{\text{TEM}}(\omega)$, which overestimates the result of Eq. (2826) by a factor $F = 8\pi / \left[ a^2 k_p^2 \ln\left(1 + \frac{4\pi}{a^2 k_p^2}\right) \right]$ that depends only on $r_w/a$. For $0.01 < r_w/a < 0.2$ the factor $F$ is of the order of 2.5-6.5.

The power emitted due to the excitation of TE modes can be calculated using similar ideas. The TE modes have an electric field such that $\mathbf{E}_{\text{TE},\mathbf{k}} = A' \mathbf{k} \times \hat{\mathbf{z}}$ with $A' = 1/\sqrt{(k_x^2 + k_y^2)\varepsilon_0 \varepsilon_h}$ and dispersion $\omega_{\text{TE},\mathbf{k}} = c_h k$. Hence, straightforward calculations show that:

$$P_{\text{rad,TE}} = \frac{\omega^4 |p_e|^2}{16\pi c_h^3 \varepsilon_0 \varepsilon_h}. \tag{27}$$

For relatively low frequencies $P_{\text{rad,TE}} \ll P_{\text{rad,TEM}}$ and hence most of the emitted power is transported by TEM waves.

It is interesting to note that the power radiated by the same electric dipole in a homogeneous dielectric with the same permittivity as the host medium is $P_{\text{rad,diel.}} = \frac{\omega^4 |p_e|^2}{12\pi c_h^3 \varepsilon_0 \varepsilon_h}$, which is only marginally larger than $P_{\text{rad,TE}}$. Moreover, the density of states of P-polarized waves in the homogeneous dielectric host is given by $D_{\text{P,diel.}}(\omega) = \omega^2/(2\pi^2 c_h^3)$ (the total density of states, including the S-polarized waves, is twice as large). Hence, it is possible to write the power emitted by the dipole in the wire medium as:

$$P_{\text{rad,WM}} = P_{\text{rad,TEM}} + P_{\text{rad,TE}}$$
$$\approx P_{\text{rad,TEM}} = P_{\text{rad,diel.}} \frac{3}{2} \frac{1}{F} \frac{D_{\text{TEM}}}{D_{\text{P,diel.}}}, \quad (28)$$

where $F$ is defined as before. Thus, the enhancement of the emitted power in the presence of the PEC nanowires is roughly proportional to the ratio of the density of states $D_{\text{TEM}} / D_{\text{P,diel.}}$. Therefore, the anomalously high density of photonic states of the uniaxial wire medium implies indeed a strong enhancement of the power radiated by the dipole.

The ratio $\frac{P_{\text{rad,TEM}}}{P_{\text{rad,diel.}}} = \frac{3}{2}\frac{1}{F}\frac{D_{\text{TEM}}}{D_{\text{P,diel.}}}$ is the Purcell factor. It is given by $\frac{3}{8}\frac{k_p^2 c_h^2}{\omega^2}\ln\left(1 + \frac{4\pi}{a^2 k_p^2}\right)$, which coincides precisely with the formula (34) of Ref. [12] derived using rather different ideas.

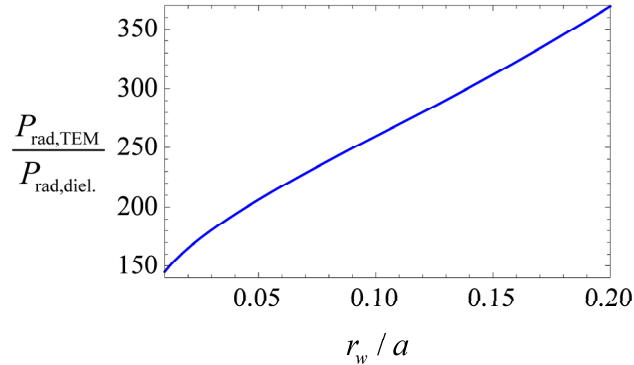

**Fig. 5.** Power transported by the TEM modes normalized to the power emitted by the electric dipole in a homogeneous dielectric with the same permittivity as the host medium, as a function of the wires radius $r_w$ calculated with Eq. (26). The host is a vacuum $\varepsilon_h = 1$, and the frequency of operation is $\omega a / c = 0.1$.

In Fig. 5 we show the dependence of the radiated power with the radius of the wires. The radiated power is larger for thicker wires, which can be understood by noting that the coupling efficiency between the point source and the nanowires is better for thick wires.

## V.  CONCLUSION

In this work, we have studied the radiation of a short horizontal electric dipole embedded in a uniaxial wire medium using an effective medium approach. It was shown that the radiation pattern of a short dipole inside a PEC wire medium corresponds to a non-diffractive beam, and thus the fields are super-collimated along the direction parallel to the nanowires. Moreover, we derived a closed analytical expression for the power radiated by the dipole relying on an eigenfunction expansion [37]. It was demonstrated that, owing to a singularity in the density of photonic states of the uniaxial wire medium, the power radiated by the dipole is strongly enhanced as compared to the power emitted by the same dipole in the host dielectric. For realistic metallic wires the isofrequency contours will become slightly hyperbolic and hence the radiated beam is expected to be slightly divergent. Finally, we note that when the length of the wire medium is finite along the $z$-direction the supercollimated beam will create a sharp near-field distribution with subwavelength features at the interface with an air region. Only the spatial harmonics with $k_\parallel < \omega/c$ can be coupled to the propagating waves in free-space, and hence the rest of the energy will stay trapped in the wire medium slab.

**Acknowledgement:** This work is supported in part by Fundação para a Ciência e a Tecnologia grant number PTDC/EEI-TEL/2764/2012. T. A. Morgado acknowledges financial support by Fundação para a Ciência e a Tecnologia (FCT/POPH) and the cofinancing of Fundo Social Europeu under the Post-Doctoral fellowship SFRH/BPD/84467/2012.